\documentclass[useAMS,usenatbib,usegraphicx]{mn2e}
\usepackage{times}
\usepackage{amsmath}
\usepackage{amssymb}

\newcommand*{\cii}{\text{[C\,\textsc{ii}]}}
\newcommand*{\nii}{\text{[N\,\textsc{ii}]}}
\newcommand*{\oiii}{\text{[O\,\textsc{iii}]}}
\newcommand*{\oi}{\text{[O\,\textsc{i}]}}
\newcommand*{\hii}{\text{H\,\textsc{ii}}}
\newcommand*{\herschel}{\emph{Herschel}}
\newcommand*{\J}[2]{\ensuremath{\text{\emph{J}} \!
                                = \! #1 \! \rightarrow\! #2}}
\newcommand*{\kms}{\text{km\,s\(^{-1}\)}}
\newcommand*{\lir}{\ensuremath{L_\text{IR}}}
\newcommand*{\lsun}{\ensuremath{\text{L}_{\sun}}}
\newcommand*{\msun}{\ensuremath{\text{M}_{\sun}}}

\newcommand*{\um}{\ensuremath{\umu \text{m}}}

\newcommand*{\smm}{\text{SMM\,J2135}}

\title[A molecular outflow in the Cosmic Eyelash]{
  \herschel{} reveals a molecular outflow in a
  \(\boldsymbol{z = 2.3}\) ULIRG}
\author[George et al.]{%
R.\,D.~George,\(^{\! 1}\) R.\,J.~Ivison,\(^{\! 1,2}\)
Ian~Smail,\(^{3}\)
A.\,M.~Swinbank,\(^{\! 3}\)
R.~Hopwood,\(^{\! 4}\) \and
F.~Stanley,\(^{\! 3}\)
B.\,M.~Swinyard,\(^{\! 5,6}\)
I.~Valtchanov\(^{7}\)
and
P.\,P.~van der Werf\(^{\, 8}\)
\vspace*{1mm}\\
\(^1\) Institute for Astronomy, University of Edinburgh,
       Royal Observatory, Blackford Hill, Edinburgh EH9 3HJ, UK\\
\(^2\) European Southern Observatory, Karl Schwarzschild Strasse 2,
        D-85748 Garching, Germany\\
\(^3\) Institute for Computational Cosmology, Durham University,
       South Road, Durham DH1 3LE, UK\\
\(^4\) Physics Department, Imperial College London,
       South Kensington Campus, London SW7 2AZ, UK\\
\(^5\) RAL Space, Rutherford Appleton Laboratory,
       Didcot OX11 0QX, UK\\
\(^6\) Department of Physics and Astronomy,
       University College London, Gower Street, London WC1E 6BT, UK\\
\(^7\) Herschel Science Centre, European Space Astronomy Centre,
       ESA, 28691 Villanueva de la Ca\~nada, Spain\\
\(^8\) Leiden Observatory, Leiden University, PO Box 9513,
       NL-2300 RA Leiden, The Netherlands
}
\date{
Submitted to MNRAS
}
\pagerange{000--000}
\usepackage[usenames,dvipsnames,svgnames,table]{xcolor}
\begin{document}

\maketitle

\begin{abstract}
We report the results from a 19-h integration with the SPIRE
Fourier Transform Spectrometer aboard the \emph{Herschel Space
  Observatory} which has revealed the presence of a molecular
outflow from the Cosmic Eyelash (SMM\,J2135\(-\)0102)
via the detection of blueshifted OH absorption.  Detections of
several fine-structure emission lines indicate low-excitation \hii{}
regions contribute strongly to the \cii{} luminosity in this \(z =
2.3\) ULIRG.  The OH feature suggests a maximum wind velocity of
\(700 \, \kms\), which is lower than the expected escape velocity
of the host dark matter halo, \({\approx} 1000 \, \kms\).  A large
fraction of the available molecular gas could thus be converted into
stars via a burst protracted by the resulting gas fountain, until an
AGN-driven outflow can eject the remaining gas.
\end{abstract}


\section{Introduction}
\label{introduction}

With only 10--30 per cent of the gas of a giant molecular cloud (GMC)
undergoing conversion into stars, and another 40 per cent of that
returning to the interstellar medium (ISM) through supernovae and
stellar winds, star formation is an inefficient process
\citep{2011ApJ...729..133M}.  The rapid dispersal of GMCs by the
combined radiation and mechanical pressure exerted by newly-formed
massive stars is responsible for this inefficiency, lowering the
intensity but extending the duration of the star formation resulting
from an initial quantity of molecular material.  This is visible in
the relatively small observed reduction in the gas available for star
formation since \(z \sim 2\) \citep[of the order of \(10 \times\);
e.g.][]{2013ARA&A..51..105C}, and in the similar decrease in the
cosmic star formation rate (SFR) in galaxies over the same period
\citep[e.g.][]{2013ApJ...770...57B, 2014MNRAS.437.3516S}.  Conversion
to stars is not the only fate of molecular gas; stellar feedback,
along with that from AGN, is also capable of expelling interstellar
material from a galaxy entirely \citep{2005ARA&A..43..769V}.

\begin{figure*}
\centering
\includegraphics[]{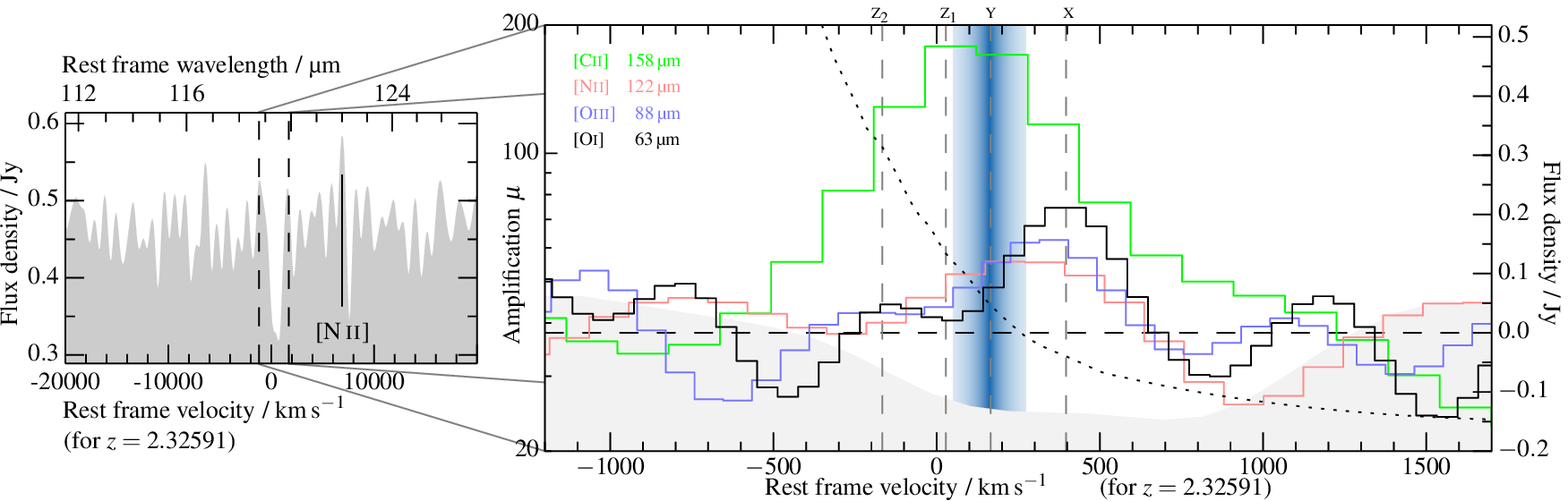}
\caption{Left: rest-frame stack centred on the OH
119.233 \um{} line at \(z = 2.32591\), the low-\(J\)
luminosity-weighted redshift derived in \citet{2011ApJ...742...11S}.
The continuum flux level is normalized to that from the spectral
energy distribution (SED) \citep{2010A&A...518L..35I}.
The vertical black dashed lines demarcate the region displayed in the
right-hand plot.  Right: zoom-in on the OH absorption line with
the rest-frame primary atomic and ionic fine-structure emission lines
in the SPIRE FTS spectral range overplotted.  The lensing
amplification as a function of velocity is shown as a dotted black
line, suggesting a higher amplification for \cii{} relative to the
other emission lines.  Note that the SPIRE FTS produces a sinc line
profile, with a width constant in frequency, and that all of the
atomic and ionic lines are spectrally unresolved.  Vertical dashed
lines are displayed at the velocities of the CO-bright clumps
identified in \citet{2013MNRAS.436.2793D}.  The vertical shaded region
corresponds to the derived median and \(1\sigma\) velocity of the OH
feature assuming two identical components and corresponds to a shift
of \({-} 240 \pm 110 \,\kms\) relative to the CO clump `X', which
contains the densest molecular gas, and which we associate with the
primary source of the \nii, \oiii{} and \oi{} emission
(see \S\ref{sec:fs_lines}).}
\label{fig:zoom}
\end{figure*}

Occurring as galaxy-scale superwinds, these outflows are observed
locally to be primarily powered by star formation, excluding those
generated in late-stage galaxy mergers, which are typically driven by
AGN \citep{2013ApJ...776...27V}.  Superwinds seem to be more common
and powerful at high redshift \citep[e.g.][]{2009ApJ...692..187W,
  2011MNRAS.418.1071B} where the majority of outflowing gas arises
from strongly star-forming, high-mass galaxies and the outflow
strength increases with SFR surface density \(\Sigma _\text{SFR}\)
\citep{2012ApJ...761...43N}.  Supernovae \citep{2009ApJ...697.2030S},
stellar winds \citep{2005ApJ...618..569M}, radiation pressure
\citep{2011ApJ...735...66M} and cosmic rays
\citep{2012MNRAS.423.2374U} may all contribute to the driving
pressure.  While star formation is expected to produce wind velocities
as high as 700 \kms{} \citep{2005ApJ...621..227M}, this may be
insufficient for the material to escape the global potential well,
leading to a `fountain': material subsequently falls back, leading to
star formation over a time-scale much greater than that of the original
burst \citep{2013MNRAS.430.1901H}.  AGN may be required for complete
expulsion of gas to the surrounding medium, since some seem able to
drive outflow velocities of over 1000\,\kms{}
\citep{2005ApJ...632..751R, 2013ApJ...775..127S}.

As with the ISM from which it originates, a superwind is expected to
comprise multiple gas phases moving with different velocities
\citep[e.g.][]{2005ApJ...632..751R, 2010A&A...518L.155F}.  Whilst
observations of X-ray and UV absorption lines have indicated ionized
and neutral atomic components \citep[e.g.][]{2012ApJ...758..135K}, the
presence of molecular gas may also be required (depending on the
time-scale for material to cycle from atomic to molecular) if future
star formation is to be reduced.  CO and OH are the most common
non-H\(_2\) molecules at the temperatures and densities involved, with
outflows inferred from observations of both species towards local
galaxies and QSOs
\citep[e.g.][]{2009ApJ...700L.104S,2010A&A...518L..41F}, indicating a
correlation between AGN luminosity and maximum outflow velocity
\citep{2013ApJ...775..127S, 2013ApJ...776...27V}, and showing that
outflows are prevalent among local ULIRGs.

OH is particularly well suited for tracing molecular gas in galactic
winds. It possesses a dipole moment \({\sim} 15 \times\) that of
CO.  Due to the high densities required to thermalize OH rotational
transitions \citep[\({\sim} 10^9 \,
\text{cm}^{-3}\);][]{1981ApJ...244L..27S}, the strongest OH line is
typically the ground-state absorption \(^{16}\text{OH} \; ^2\Pi_{3/2}
\; \J{\frac{3}{2}}{\frac{5}{2}} \), split by lambda-doubling into
rest-frame 119.233 and 119.441\,\um{} transitions.  Back-lit molecular
gas provides a prime opportunity to observe these transitions, with a
blueshift expected from the radial movement of an outflow.

Despite the likely higher prevalence of superwinds at high redshift,
and the potential advantages of observing OH, the far-infrared (FIR)
rest-frame wavelengths of these transitions necessitate the use of
space-based facilities. The consequential small apertures have meant
that observations have been restricted to local galaxies
\citep[e.g.][]{2013ApJ...775..127S, 2013ApJ...776...27V}.  Previous
studies of high-redshift outflows have generally utilized UV
lines \citep[e.g.][]{2010ApJ...717..289S,2012ApJ...760..127M}, which
are generated within ionized material and thus insensitive to any
molecular phase, critical to determination of whether superwinds or
\emph{in situ} star formation dominates the reduction in interstellar
molecular gas and SFR density at early epochs.

Gravitational lensing provides a means to circumvent these
limitations.  The galaxy, SMM\,J2135\(-\)0102, hereafter \smm,
was identified in the field of a
massive, lensing cluster via ground-based 870-\um{} imaging
\citep{2010Natur.464..733S}.  Subsequent imaging with high spatial
resolution constrained the lensing model, showing that \smm{} is
resolved into an \({\sim} 5 \, \text{arcsec}\) long arc, mirrored
about the critical curve, with a total magnification
\(\mu = 37.5 \pm 4.5\) \citep{2011ApJ...742...11S}.
CO measurements of its velocity field
reveal a disc with rotation speed \(320 \, \kms\), dominated by
several star-forming clumps, each of radius 100--200\,pc, distributed
across a radius of \({\sim} 2 \, \text{kpc}\).

Here, we describe the discovery of a molecular outflow from \smm, as
betrayed by the detection of the 119\,\um\ OH transition in absorption
during a number of long integrations with the \emph{Herschel Space
Observatory}.  Throughout, we use \emph{Planck} 2013 cosmology
\citep{2013arXiv1303.5076P} with \(H_0 = 67.8 \, \kms \,
\text{Mpc}^{-1}\), \(\Omega_\text{m} = 0.307\) and \(\Omega_\Lambda =
0.693\).

\section{Observations}
\label{sec:observations}

Due to its high magnification, correspondingly high FIR flux densities
and previous in-depth studies \citep[e.g.][]{2010A&A...518L..35I,
  2011MNRAS.410.1687D, 2013MNRAS.436.2793D}, \smm{} was included in
our programme of deep \herschel{} \citep{2010A&A...518L...1P} SPIRE
FTS \citep{2010A&A...518L...3G} observations
\citep[e.g.][George et. al. in preparation]{2011MNRAS.415.3473V,
2013MNRAS.436L..99G}.

We initially observed \smm{} as part of \textsc{ot1\_rivison\_1}: a
3.8-h integration using the central detectors of the SPIRE arrays,
comprising 200 scans with the FTS mirror.  The data were processed
using the \herschel{} data processing pipeline
\citep{2010SPIE.7731E..99F} within the Herschel Interactive
Processing Environment \citep{2010ASPC..434..139O} v11.  The local
background was estimated as the mean spectrum observed by the rings of
detectors surrounding the central detector.

The resulting spectrum covering \(\lambda_\text{obs} =
\text{194--671\,\um}\) provided strong detections of several
fine-structure lines, along with a tentative indication of OH
absorption.  We then obtained five additional 3.8-h integrations as
part of \textsc{ot2\_rivison\_2}, processing these using the same
method as the first observation, in order that the spectra may be
co-added.  We rejected one of these, due to inconsistencies across
several wavelength ranges likely caused by poor background estimation
as a result of one of the surrounding detectors producing corrupt
data.  The combined integration time of the useful data is therefore
19\,h.

\section{Results and discussion}
\label{sec:results}

The continuum-subtracted rest-frame stack of our observations provides
a significant detection of \cii, as previously identified by
\citet{2010A&A...518L..35I}, alongside new detections of
\nii\,122\,\um, \oiii\,88\,\um{} and \oi\,63\,\um.  The previous,
tentative indication of OH absorption is confirmed.

The low angular resolution of the SPIRE FTS is unable to spatially
resolve individual kinematic components within \smm, instead producing
a galaxy-integrated spectrum.  The velocity separation of the clumps
partially alleviates this complication, however, allowing a comparison
of the detected lines with the previous CO measurements.  The stack at
119.233\,\um{} is displayed in Fig.~\ref{fig:zoom}, along with a
comparison of the OH absorption with the rest-frame fine-structure
emission lines.

\subsection{Atomic and ionic fine-structure lines}
\label{sec:fs_lines}

\begin{table}
\centering
\begin{tabular}{
  l @{\hspace{0.3em}} r @{\hspace{1em}}
  r @{\hspace{0.2em}} c @{\hspace{0.2em}} l @{\hspace{1em}}
  r @{\hspace{0.2em}} c @{\hspace{0.2em}} l @{\hspace{2em}}
  r @{\hspace{0.2em}} c @{\hspace{0.2em}} l @{\hspace{0em}}
  r @{\hspace{0.2em}} c @{\hspace{0.2em}} l
}
\hline
\hline
\noalign{\vspace{0.4ex}}
\multicolumn{2}{c}{\hspace{-1em}Transition}
& \multicolumn{3}{c}{\hspace{-1em}Velocity\(^{\,a}\)}
& \multicolumn{3}{c}{\hspace{-1.2em}Flux}
& \multicolumn{3}{c}{\hspace{-1.5em}Amplification}
& \multicolumn{3}{c}{\hspace{-0.5em}Luminosity }\\
\multicolumn{2}{c}{\hspace{-1em}(\um)}
& \multicolumn{3}{c}{\hspace{-1em}(\kms)}
& \multicolumn{3}{c}{\hspace{-0.5em}(Jy\,\kms)}
& \multicolumn{3}{c}{}
& \multicolumn{3}{c}{\hspace{-0.5em}(\(10^9\,\lsun\))}\\
\noalign{\vspace{0.3ex}}
\hline
\noalign{\vspace{0.3ex}}
\multicolumn{2}{l}{Low-\(J\) CO} & & 0 & & & & & 37.5 & $\pm$ &  4.5 & & & \\
\cii  & 158 & 100 & $\pm$ & 140 & 280    & $\pm$ & 80 & 50 & $\pm$ & 20 & 1.2    & $\pm$ & 0.6\\
\nii  & 122 & 290 & $\pm$ & 100 & 60     & $\pm$ & 30 & 36 & $\pm$ & 12 & 0.5    & $\pm$ & 0.3\\
\oiii & 88  & 330 & $\pm$ & 130 & 50     & $\pm$ & 40 & 35 & $\pm$ & 17 & 0.6    & $\pm$ & 0.5\\
\oi   & 63  & 400 & $\pm$ & 90  & 50     & $\pm$ & 30 & 33 & $\pm$ & 12 & 0.8    & $\pm$ & 0.6\\
OH    & 119 & 160 & $\pm$ & 110 & $-180$ & $\pm$ & 60 & 33 & $\pm$ & 5  & $-1.6$ & $\pm$ & 0.6\\
\hline
\end{tabular}
\begin{tabular}{l@{\extracolsep{0.3em}}p{27em}}
\(^a\) & Velocities are relative to the low-\(J\) CO
         luminosity-weighted redshift of \(z = 2.32591\)
         \citep{2011ApJ...742...11S}.
\end{tabular}
\caption{FIR line properties from fitting sinc profiles (as
appropriate for an FTS), with errors determined via jackknife
data sets.  All fine-structure lines are unresolved (1.44\,GHz FWHM),
with luminosities calculated using the lensing amplification factors
derived for their observed velocities
\citep{2011ApJ...742...11S}.  The OH properties were derived by
integrating over all flux less than the continuum level and assuming
two identical components, and assume an amplification corresponding to
a source velocity of \(400 \pm 50 \, \kms\).}
\label{tab:fs_lines}
\end{table}

\begin{figure}
\centering
\includegraphics[]{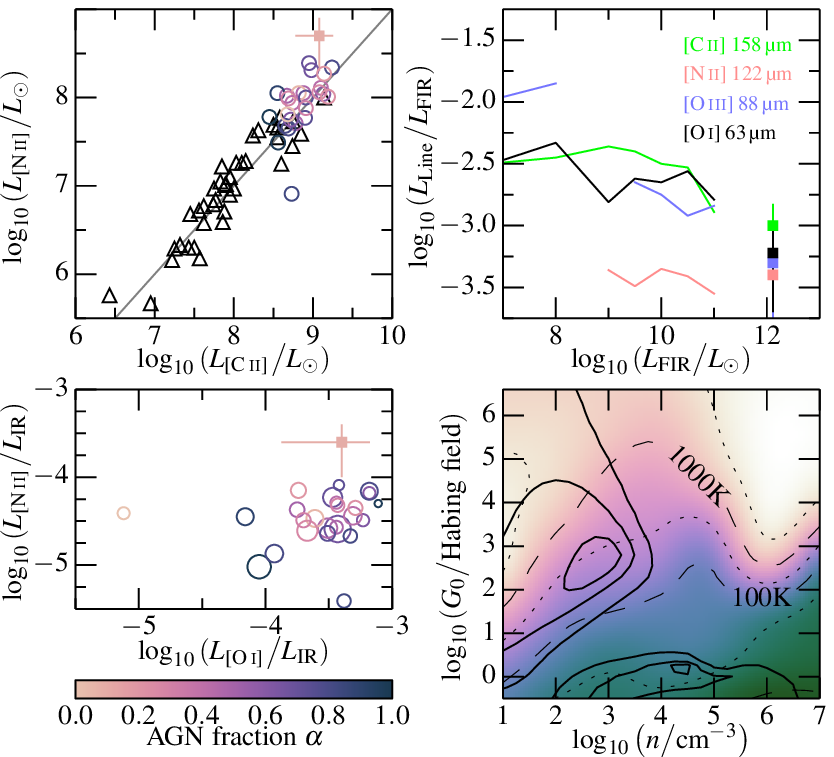}
\caption{Properties of the atomic and ionic fine-structure lines.
Upper left: comparison of line luminosities in \smm{} (filled
square) with local ULIRGs \citep[coloured open circles, colour map as
in lower left plot;][]{2013ApJ...776...38F} and local galaxies of
various types
\citep[black open triangles;][]{2008ApJS..178..280B}.  The line
\(L_\nii = 0.1 L_\cii\) is shown in grey.  Upper right: mean
line-to-FIR luminosity ratios for several lines in the
\citet{2008ApJS..178..280B} sample.  \smm{} values are denoted by
squares.  Lower left: line-to-IR luminosity ratios in
\citet{2013ApJ...776...38F} and \smm{} (square).  Colour map displays
the fractional contribution of AGN to
\(L_\text{bol}\), with marker area proportional to \(L_\text{bol}\).
Lower right: estimated \(\chi^2\) contours for CO clump, `X'
\citep{2011MNRAS.410.1687D}, of \smm{} (red) overlaid on the PDR
surface temperature model from \citet{2006ApJ...644..283K},
suggesting \(n\) and \(G_0\) values of \(10^{2 \text{--} 3}\) (see
text for caveats).}
\label{fig:line_luminosities}
\end{figure}

To determine the properties of the observed FIR lines, we fitted each
with a sinc profile, as appropriate for an FTS spectrum
\citep{2014MNRAS.440.3658S}.  The resulting line measurements are detailed in
Table~\ref{tab:fs_lines}, with errors conservatively estimated as the
maximum obtained via the use of two methods.

The first method involves applying the jackknife technique to the
individual FTS scans (200 for each observation).  We generate 20
subsets of 190 scans, neglecting 10 consecutive scans at a time, fit a
sinc profile to each line and determine the variance of the fit
parameters from the 20 subsets.  The second method requires cutting a
region around each spectral line, estimating the variance and standard
deviation of the variance of that region.  Best-fitting and
\(1 \sigma\) line parameters are then calculated such that subtracting
a sinc profile with those values returns residuals with the estimated
and \(1 \sigma\) variances.

Intriguingly, several of the fine-structure line velocities appear to
be offset from the literature redshift, derived from
intensity-weighted low-\(J\) CO emission, used to define the zero
velocity in Fig.~\ref{fig:zoom} and Table~\ref{tab:fs_lines}, with the
\hii{} region lines at higher velocities than \cii\,158\,\um, a
photodissociation region (PDR) line.
We observe the peak \oi\,63\,\um{} emission at \(400 \pm 90 \, \kms\),
an unexpected value given the status of this transition as
a primary coolant of dense PDRs, and hence typically colocated with
\cii.  The CO component, `X', was identified in
\citet{2011MNRAS.410.1687D, 2013MNRAS.436.2793D} to be the most active
site of massive star formation in \smm, with a
velocity of \({\sim} 400 \, \kms\). The \nii{} and \oiii{} transitions
require the presence of an ionizing source, and considering the large
uncertainties on the velocity offsets, we associate these lines with
this CO component, acknowledging that the molecular and ionized gas
may not be completely cospatial.  The material traced by these lines
appears to be very different from that observed in CO and with the
Submillimeter Array (SMA), with individual line profiles displaying no
structure, although the spectral resolution of the FTS may smear out
finer details.

In contrast to several other high-redshift starbursts
\citep[e.g.][]{2011ApJ...740L..29F, 2011MNRAS.415.3473V}, we observe a
line-to-continuum flux deficit similar to that seen in local ULIRGs
\citep{2003ApJ...594..758L}.  Most of the FIR lines have
luminosity-to-\(L_\text{FIR}\) ratios slightly below the mean of those
of the local \citet{2008ApJS..178..280B} sample
(Fig.~\ref{fig:line_luminosities}), consisting of low-redshift normal
star-forming galaxies, starbursts and AGN, compared to which \smm{}
possesses an order-of-magnitude higher FIR luminosity.  A comparison
(Fig.~\ref{fig:line_luminosities}) to the \(z \sim 0.1\) ULIRG sample
of \citet{2013ApJ...776...38F} reveals comparable line luminosities
and \(L_\text{line}\)-to-\(L_\text{IR}\) ratios in \smm, with the
exception of \nii{} which is stronger in \smm{} than for any galaxy in
that study.

The \cii{} results in particular are interesting, with the derived
luminosity a factor of 4 lower than estimated in
\citet{2010A&A...518L..35I}. The earlier value was derived assuming a
smaller lensing amplification (32.5) and using a Gaussian rather than
a sinc fit, from a much shallower spectrum calibrated with the
asteroid Vesta (Uranus is now used).  Our estimate gives a line
luminosity-to-\(L_\text{IR}\) ratio of 0.06 per cent.  With a
\cii/\nii{} ratio of 2.5, similar to that tentatively observed in
LAE-2 by \citet[and much lower than its companion submillimetre galaxy
(SMG) in that paper]{2014ApJ...782L..17D}, \hii{} regions
may make a major contribution to the \cii{} luminosity.  An
intriguing, if perhaps unlikely scenario is a strong contribution to
the \cii{} emission from the outflow itself, which would suggest a
lower amplification (due to increased source velocity) and hence
higher luminosity.

With the \oi{} emission appearing to arise from within CO clump, `X',
we fit several observed parameters for this clump to the PDR models
presented in \citet{2006ApJ...644..283K}.  We assume an FIR luminosity
for this clump, \(L ^\text{X} _\text{FIR} = L ^\text{tot} _\text{FIR}
\times \text{SFR}^\text{X} / \text{SFR}^\text{tot}\), and that 50 per
cent of \cii{} emission arises from the \oi-producing PDRs within this
clump (a value likely overestimated should the observed \cii--\oi{}
velocity offset prove genuine and \hii{} regions contribute strongly
to the \cii{} luminosity).

The [C\,\textsc{i}]\,370\,\um/[C\,\textsc{i}]\,609\,\um{} ratio from
\citet{2011MNRAS.410.1687D} should additionally probe the deeper
regions \citep[\(A_\text{V} \sim 2 \text{--}
4\);][]{2006ApJ...644..283K}.  As indicated in the lower-right panel
of Fig.~\ref{fig:line_luminosities}, these parameters indicate a gas
density \(n \sim 10^{2 \text{--} 3} \, \text{cm} ^{-3}\) and incident
FUV flux \(G_0 \sim 10^{2 \text{--} 3}\) Habing fields
(Fig.~\ref{fig:line_luminosities}) within CO clump `X', consistent
with \citet{2013MNRAS.436.2793D}.  The above values both increase as
the \cii/\oi{} ratio drops, with \citet{2013MNRAS.436.2793D}
estimating \(n \sim 10^3 \, \text{cm} ^{-3}\) for the molecular gas
within clump `X'.

With dissimilar photon energies required for production of the species
involved, but comparable critical densities, the relative luminosity
\(L_\oiii / L_\nii\) is a sensitive diagnostic of the interstellar
radiation field hardness, little affected by gas density.  We observe
\(L_\oiii / L_\nii = 1.2 \pm 1.2\) in \smm, with the models presented
in \citet{2011ApJ...740L..29F} thus indicating production within an
\hii{} region surrounding a starburst with an effective temperature of
35\,000\,K or within a narrow-line region with ionization parameter
\({\sim} {-} 3.5\) (both models assume solar metallicity).  To
investigate the likelihood of an AGN-associated narrow-line region,
we estimate the maximum possible AGN luminosity from the SED.  We use
the IR SED fitting method introduced in \citet{2011MNRAS.414.1082M},
and extended in \citet{2013A&A...549A..59D} to fit to the available
photometry from MIR to sub-mm wavelengths.  This SED fitting method
allows the decomposition of the AGN and host contribution to the
galaxy's IR emission.  The best-fitting result requires no AGN
contribution, attributing the whole of the IR emission to the host
galaxy.  To place a constraint on the AGN IR emission, we assume an
upper limit that corresponds to 10 per cent of the resulting total
emission from the best fit.  This, along with the higher total
\(L_\nii\) and \(L_\cii\) and low \(G_0\) inferred above, is
consistent with a situation in which SMGs possess more extended star
formation than local ULIRGs, generated by large, low-ionization \hii{}
regions
\citep[e.g.][]{2008MNRAS.385..893B, 2009ApJ...699.1610H,
  2009ApJ...699..667M, 2011MNRAS.412.1913I, 2011ApJ...742...11S}.

\subsection{OH absorption}

\begin{figure}
\centering
\includegraphics[]{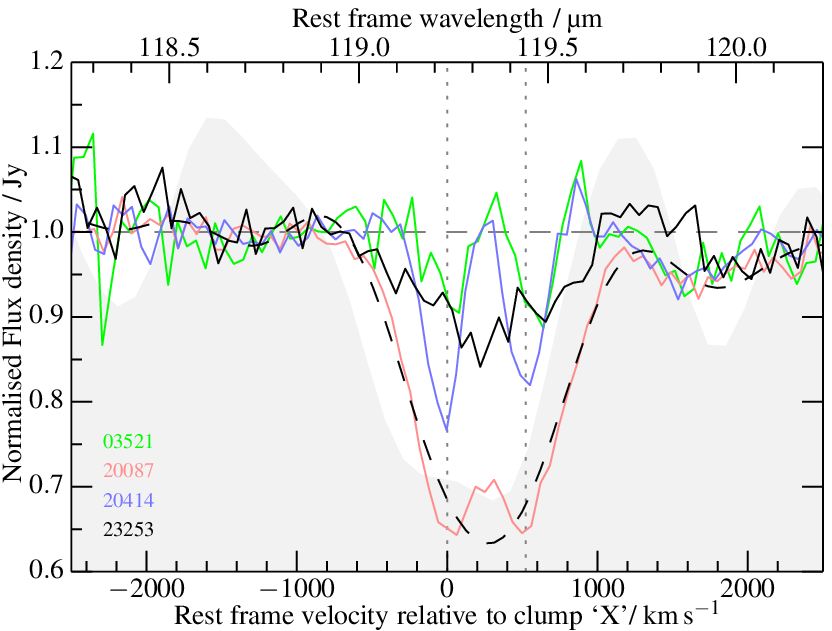}
\caption{Normalized OH profile of \smm{} compared to those of the four
starburst-dominated (AGN contribution \(<25\) per cent of \lir)
sources in the local sample of \citet{2013ApJ...775..127S}, observed
with \herschel/PACS \citep{2010A&A...518L...2P}.  The low maximum
and median blueshift velocities of these (including \smm) are typical
of systems with low AGN contributions (see
Fig.~\ref{fig:sfr_lagn_vmax}).  Vertical dotted lines denote the
rest-frame velocities of the OH doublet.  The velocity scale in this
figure is set relative to that of the assumed driving source (such
that the luminosity-weighted low-\(J\) CO redshift for the system as a
whole corresponds to \(-400 \, \kms\)).   Note that the SPIRE FTS
produces an FWHM of \({\sim} 570 \, \kms\) at this observed frequency,
with the PACS spectrometer producing an FWHM of \({<} 290 \, \kms\)
for the displayed observations.  Shown as a black dashed line is the
20087 spectrum deconvolved with the PACS spectral profile and then
convolved with the SPIRE spectral profile, producing a very similar
profile to that observed in \smm.}
\label{fig:oh_starburst_comparison}
\end{figure}

Due to the relatively poor spectral resolution of the SPIRE FTS, we do
not see a readily-identifiable P-Cygni profile in the 119\,\um{} OH
feature and as such we fitted it with two identical components
constrained to have the rest-frame frequency separation of the line
doublet, suggesting a velocity offset of \(160 \pm 110 \,\kms\)
(see Fig.~\ref{fig:zoom} and Table~\ref{tab:fs_lines})
with respect to the low-\(J\) CO luminosity-weighted redshift of
\(z = 2.32591\) \citep{2011ApJ...742...11S}.  Requiring
strong active star formation from dense material as the primary
driver, as with the \nii, \oiii{} and \oi{} lines, we associate the OH
source with CO clump, `X' \citep{2013MNRAS.436.2793D}, at \(v \sim 400
\, \kms\) corresponding to a mean velocity of the OH of \({-}240 \pm
110 \, \kms\).  \citet{2005ApJ...632..751R} require \(v < -50 \,
\kms\) for classification as an outflowing wind.  As such, this
\(-180 \pm 40 \, \text{Jy} \, \kms\) feature represents the highest
redshift observation of outflowing molecular gas.

\begin{figure*}
\centering
\includegraphics[]{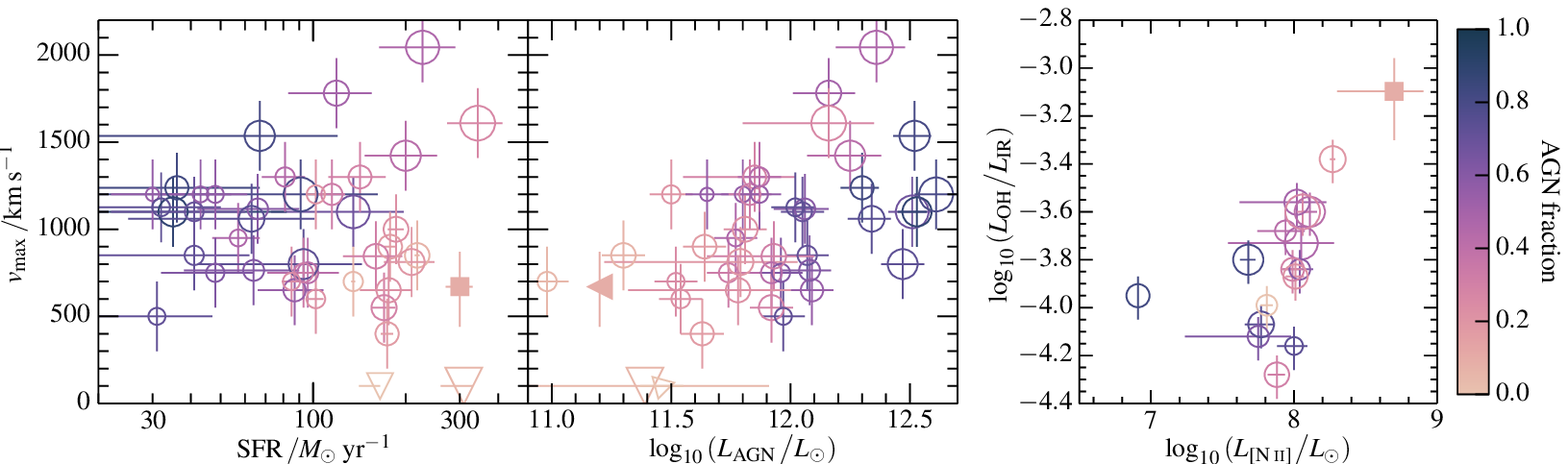}
\caption{Left: comparison of \smm{} to local ULIRGs from
\citet{2013ApJ...775..127S} and \citet{2013ApJ...776...27V}.  Marker
colour denotes the fractional AGN contribution to \(L_\text{bol}\)
(see left-hand colour bar), with enclosed marker area proportional
to \(L_\text{bol}\).  \smm{} is displayed with a filled marker, and
appears typical of sources with little AGN contribution within that
sample, with \(L_\text{AGN}\) the primary determinant of
\(v_\text{max}\).  The AGN luminosity indicated for \smm{} is
derived from the estimated maximum 10 per cent contribution to
\(L_\text{IR}\) (see \S\ref{sec:fs_lines}).  Right: comparison with
the \citet{2013ApJ...775..127S} and \citet{2013ApJ...776...38F}
samples.  The much lower OH-to-IR luminosity ratios of the ULIRGs of
that sample likely indicates higher OH optical thickness or larger
OH covering fractions.}
\label{fig:sfr_lagn_vmax}
\end{figure*}

No other OH transitions are observed, consistent with the faint
79\,\um{} cross-ladder transition observations in
\citet{2013ApJ...775..127S}.  Multiple transitions have been detected
in the nearby quasar, Mrk\,231 \citep{2014A&A...561A..27G}, with
several showing absorption fluxes comparable to that of the 119\,\um{}
doublet, suggesting that the 119\,\um{} lines are optically thick in
that system, or that the radiatively excited upper levels are more
highly populated.  The large full width at half maximum (FWHM) of the
FTS spectral profile makes it
difficult to determine the underlying line structure; however, a
comparison to the velocities of the four clumps indicates that it is
likely that the outflow does not have a covering factor of unity.

In Fig.~\ref{fig:oh_starburst_comparison}, we compare the 119\,\um{}
OH absorption feature of our stacked spectra, to that of
starburst-dominated sources from the \citet{2013ApJ...775..127S}
sample.  The normalized, integrated, absorbed flux,
\(-380 \, \text{Jy} \, \kms\), is at the high end of those in that
sample, including both starburst- and AGN-dominated sources.

Following \citet{2013ApJ...775..127S}, we define \(v^\text{obs}
_\text{max}\) as the velocity at which a B-spline fitted to the blue
absorption wing crosses the continuum flux density.  Correcting for
turbulence (100\,\kms; as used in \citealp{2013ApJ...775..127S}), and
instrumental spectral resolution (1.2\,GHz) in quadrature, we obtain
an estimate for the maximum velocity of outflowing molecular gas:
\(v ^\text{mol} _\text{max} = -670 \, \kms\) for
\(v_\text{source} = 400 \, \kms\), with \citet{2013ApJ...775..127S}
estimating the uncertainty on \(v^\text{mol}_\text{max}\) obtained in
this method as \(\pm 200\,\kms\).  This similarity to the maximum
predicted for starburst-driven outflows
\citep[\({\sim} 700\,\kms\);][]{2005ApJ...621..227M} suggests highly
efficient momentum transfer within \smm.

It is additionally instructive to compare this result with outflows
from high-redshift galaxies determined from UV absorption lines.
Winds from SMGs should likely possess qualities of both these, and of
outflows from local ULIRGs.  Fig.~\ref{fig:sfr_lagn_vmax} shows our
detection in relation to those presented in
\citet{2013ApJ...775..127S} and \citet{2013ApJ...776...27V}, with
\smm{} occupying the lower-right corner of the
\(v ^\text{mol} _\text{max}\)--SFR plot, typical of the sources with
low fractional AGN contributions to \(L_\text{bol}\).  A correlation
of \(v^\text{UV}_\text{max}\) with SFR surface density
\(\Sigma_\text{SFR}\) in such high-redshift galaxies has been
identified \citep{2010AJ....140..445C, 2012ApJ...758..135K,
  2012ApJ...761...43N}.  \(\Sigma_\text{SFR}\) is typically found in
high-redshift SMGs to be between that in low-redshift ULIRGs and
high-redshift `normal' galaxies, suggesting comparatively high
expected \(v ^\text{UV} _\text{max}\) within \smm{} and other SMGs;
however, the observed ionized components may have velocities
systematically different from that of the molecular phases, with
\citet{2011MNRAS.418.1071B}, \citet{2012ApJ...758..135K} and
\citet{2012ApJ...760..127M} finding \(v ^\text{UV} _\text{max}\)
similar to \(v^\text{mol}_\text{max}\) for \smm{} within galaxies of
substantially lower \(\Sigma_\text{SFR}\).

With an inclination, \(60 \pm 8^\circ\), determined by
\citet{2011ApJ...742...11S} for \smm, we now consider the solid angle
into which molecular gas may be ejected.  An angle of \(140 ^{\circ}\)
perpendicular to the disc is suggested by the ULIRG samples of
\citet{2005ApJ...632..751R}, \citet{2013ApJ...775..127S} and
\citet{2013ApJ...776...27V} in which 2/3 of the OH detections show
outflows, consistent with the OH detection in \smm.  More collimated
winds are however advocated by the lower detection rates of
blueshifted optical or UV absorption-line features in sources of lower
SFR at both low \citep{2005ApJ...632..751R} and high redshift
\citep{2012ApJ...758..135K, 2012ApJ...760..127M}.  Such a disparity
could exist, should there be a difference in wind structure between
those driven by AGN and starbursts, or those generated in the
disturbed morphologies of late-stage mergers.  Our measurements for
\smm{} are consistent with the above literature, suggesting that in
starburst-dominated sources it is likely that the SFR or
\(\Sigma_\text{SFR}\) are primary determinants of the opening angle,
with AGN contributing particularly to \(v_\text{max}\)
(Fig.~\ref{fig:sfr_lagn_vmax}).  A systematic difference in opening
angle may however be difficult to disentangle from both the wind
velocity and occurrence rate as a function of various host galaxy
properties, in particular \(\Sigma_\text{SFR}\), particularly if
observational constraints restrict classification to those of the
highest absorbed flux or \(v_\text{max}\).

Recent simulations have shown starburst-powered outflow rates of up to
\(1000 \, \msun \, \text{yr}^{-1}\) \citep{2013MNRAS.433...78H},
consistent with mass-loading factors, \(\eta = \dot{M}_\text{outflow}
/ \text{SFR}\), which locally are of the order of a few and -- as with
\(v_\text{max}\) -- exhibit a positive correlation with AGN luminosity
\citep{2010A&A...518L.155F, 2014A&A...562A..21C, 2014A&A...561A..27G}.
At higher redshift, high-mass starbursting galaxies (of which SMGs are
an important sub-population) may dominate the global gas mass returned
to the IGM \citep{2012ApJ...761...43N}.

An estimate of the mass-outflow rate from \smm{} can be determined
from the column density of ground-state OH molecules via a physical
model, such as that presented in \citet{2014A&A...561A..27G}, a
modified version of which is now detailed.

Radially outflowing material occurs within a region surrounding a
driving source at a radius \({r > r' > R}\), such that the OH column
density \(N_\text{OH}(\Omega) = \int^R _r n_\text{OH}(r',\Omega) \,\,
\text{d} r'\).  Mass conservation then requires that
\(n_\text{OH}(r',\Omega) \, r'^2 \, v(r',\Omega)\) (where \(v(r')\) is
the OH velocity at radius \(r'\)) is constant across \(r'\) at any
\(\Omega\), leading to the mass outflow rate
\begin{equation}
  \dot{M} = m_{\text{H}_2} \, X_\text{OH} ^{-1}
  \, n_\text{OH}(r) \, r^2 \, v(r)
  \int w(\Omega) \,\, \text{d} \Omega,
\end{equation}
where \(m_\text{H}\) is the mass of a hydrogen atom, \(X_\text{OH} =
n_\text{OH} / n_{\text{H}_2}\) and the angular dependence has been
moved to a factor \(w(\Omega)\).  The mass conservation at \(r\) can
be expressed as a function of the column density \(N_\text{OH}\) via
the use of a simple linear velocity model \( v(r') = v + \frac{V -
  v}{R- r} (r' - r) \), where \(v = v(r)\) and \(V = v(R)\), resulting
in
\begin{equation}
  n_\text{OH} \, r ^2 \, v \! = \!
  N_\text{OH}
  \frac{R v \! - \! r V}{R \! - \! r} \! \!
  \left[
    \frac{1}{r}
    \! - \!
    \frac{1}{R}
    \! + \!
    \frac{V \! - \! v}{R v \! - \! r V}
    \, \text{ln} \! \left( \! \frac{V r} {v R} \! \right)
  \! \right]^{-1}. \hspace{-1em}
\end{equation}
To complete this model we derive an estimate of the column density
from the observed absorption:
\begin{equation}
\frac{N_\text{OH}}{\text{cm}^{-2}} = 6.7 \times 10^{10}
\left( \! \frac{T_\text{ex}}{\text{K}} \! \right)
\frac{\text{exp} \! \left( \! \frac{130 \, \text{K}}
                                   {T_\text{ex}} \! \right)}
     {\text{exp} \! \left( \! \frac{121 \, \text{K}}
                                   {T_\text{ex}} \! \right) - 1}
\left( \! \frac{\int \tau(v)\, \text{d}v}
               {\text{km} \, \text{s}^{-1}} \! \right),
\end{equation}
where \(T_\text{ex}\) is the excitation temperature of the
outflowing gas.

For physically motivated model parameter values: \(X_\text{OH} = 5
\times 10 ^{-6}\) \citep{2002ApJ...576L..77G}, \(\int w(\Omega) \,
\text{d} \Omega = \frac{2}{3} \times 4 \upi\), \(r = 100 \, \text{pc}\)
(motivated by the size of components `X' and `Y';
\citealp{2011ApJ...742...11S}), \(R = 200 \, \text{pc}\), \(v(r) = v =
670 \, \kms\), \(v(R) = V = 100 \, \kms\), \(T_\text{ex} = 200 \,
\text{K}\) \citep{2013MNRAS.436.2793D}, \(\int^{-100} _{-670} \tau(v)
\, \text{d} v = 2000 \, \kms\), the 119 \um{} absorption suggests a
mass-outflow rate of \(120 \, \msun \, \text{yr}^{-1}\) and hence a
mass loading \({\approx} 2\) for the \(70 \, \msun \, \text{yr}^{-1}\)
SFR of CO clump `X' \citep{2011ApJ...742...11S}.  We note that the
above parameters are determined only poorly and mass-outflow rates
estimated from this single transition are likely to be extremely
uncertain.  Particularly insecure are the estimates of temperature,
integrated optical depth and \(X_\text{OH}\), the latter determined
from local GMCs, whereas this outflow may be launched from a region
dominated by atomic gas, and the outflow likely contains multiple
phases regardless.

The estimated mass-outflow rate contributes sub-dominantly to the
total gas-reduction rate, which is primarily by star formation (\(300
\, \msun \, \text{yr}^{-1}\), estimated from the rest-frame
8--1000\,\um{} luminosity of \citealt{2010A&A...518L..35I}, corrected
for the more recent lensing amplification
\citealp{2011ApJ...742...11S} and using the relation of
\citealt{2011ApJ...737...67M}).  This suggests that the \(2.5 \times
10 ^{10} \, \msun\) of molecular gas within \smm{}
\citep{2013MNRAS.436.2793D}, will be exhausted on a time-scale of
\(\approx 60 \, \text{Myr}\), primarily through star formation, but
as we note above, the calculated mass-outflow rate is extremely
uncertain.  High-redshift SMGs are conjectured to have duty cycles of
\(\approx 100 \, \text{Myr}\) \citep[e.g.][]{2012MNRAS.421..284H,
  2013MNRAS.429.3047B}, comparable to that estimated here for
molecular gas removal, suggesting little conversion of atomic to
molecular gas, or gas input from the IGM.

The velocity required for outflowing gas to become unbound or to join
the hot intracluster medium cannot be determined reliably without
knowledge of the dark matter halo mass and the location of \smm{}
within that halo. However, utilizing the estimated stellar mass,
\(M_\star = 3 \times 10 ^{10} \, \msun\) \citep{2010Natur.464..733S},
alongside table 7 from \citet{2010ApJ...710..903M}, suggests
\(M_\text{DM}(r_\text{vir}) = 2 \times 10^{12} \, \msun\).  This value
neglects potential SMG clustering \citep[e.g.][]{2012MNRAS.421..284H}
so may underestimate the true halo mass.  The escape velocity from a
dark matter halo with a Navarro--Frenk--White profile
\citep{1997ApJ...490..493N} with this virial mass, a virial radius of
100\,kpc and a concentration parameter \(c = 10\), is \({\sim} 1000
\, \kms\), somewhat higher than the \(v_\text{max}\) calculated above
for \smm, supporting the suggestion that AGN are required for full
expulsion of interstellar material from massive
galaxies.  Star-formation-powered outflows such as the one we have
found in \smm{} are likely unable to escape the dark matter halo and
may merely serve to maintain the elevated SFR of SMGs though a
fountain mechanism.

\section{Conclusions}
\label{conclusions}

We present the first detection of OH absorption in a high-redshift
galaxy, exploiting the SPIRE FTS aboard \herschel, and interpret
this as a starburst-driven molecular outflow.

We additionally detect emission lines of \cii\,158\,\um,
\nii\,122\,\um, \oiii\,88\,\um{} and \oi\,63\,\um, finding line
luminosities similar to local ULIRGs and indicative of a significant
contribution to the \cii{} luminosity from \hii{} regions.  PDR models
indicate \(n\) and \(G_0\) values of \(10^{2 \text{--} 3}\)
\(\text{cm}^{-3}\) and \(10^{2 \text{--} 3}\) Habing fields,
respectively.

The 119-\um{} OH absorption is blueshifted by \(240 \pm 110 \, \kms\)
relative to the star-forming clump, `X', pinpointed previously by
low-\(J\) CO observations, and which we associate with the observed
\nii{} 122-\um{} and \oiii{} 88-\um{} emission.  This molecular
component to a starburst-driven wind, possesses a maximum velocity of
\(700 \pm 200 \, \kms\), smaller than the escape velocity expected
for the dark matter halo.  A large fraction of the available
molecular gas could thus be converted into stars via a burst
protracted by the resulting gas fountain, until an AGN-driven outflow
can eject the remaining gas.

\section*{Acknowledgements}

RDG acknowledges support from the Science and Technology Facilities
Council.  RJI acknowledges support from the European Research Council
(ERC) in the form of Advanced Grant, \textsc{cosmicism}.  IRS
acknowledges support from STFC (ST/I001573/1), the ERC Advanced Grant
\textsc{dustygal}, and a Royal Society/Wolfson Merit Award.
\herschel{} was an ESA space observatory with science instruments
provided by European-led Principal Investigator consortia and with
important participation from NASA.

\bibliography{outflow}

\bsp

\end{document}